\documentclass[%
 reprint,
groupedaddress,
amsmath,
amssymb,
aps,
prd,
twocolumn,preprintnumbers,floatfix,nofootinbib]{revtex4-2}

\usepackage[pdftex]{graphicx}
\usepackage{dcolumn}
\usepackage{bm}
\usepackage[hidelinks]{hyperref}

\newcommand{\be}{\begin{equation}}
\newcommand{\ee}{\end{equation}}
\newcommand{\bea}{\begin{eqnarray}}
\newcommand{\eea}{\end{eqnarray}}

\usepackage{mathrsfs}
\usepackage{amsmath, amsthm, amssymb}
\usepackage{color}
\usepackage{epstopdf}
\usepackage{amssymb}
\usepackage{verbatim}
\usepackage{enumerate}
\usepackage[caption = false]{subfig}
\usepackage[normalem]{ulem}  
\usepackage{cancel}

\usepackage{epsfig}
\usepackage{cleveref}

\newcommand{\apjl}{ApJLett}		

\newcommand{\physrep}{Phys.~Rep.}   

\begin{document}

\title{Yukawa-Screened Bose-Star Condensation}

\author{Jiajun Chen}
\email{chenjiajun@swu.edu.cn}
\affiliation{School of Physical Science and Technology, Southwest University, Chongqing 400715, China}

\date{\today}

\begin{abstract}
We study Bose-star formation in a Yukawa-Schr\"odinger-Poisson (YSP) system. A finite interaction range suppresses the infrared kinetic relaxation responsible for Bose-star condensation, modifying both the equilibrium Bose-star structure and the condensation timescale. We derive a screened kinetic condensation formula in which the ordinary gravitational Coulomb logarithm is replaced by a finite Yukawa transport logarithm. Static YSP solutions show that Yukawa screening broadens the Bose-star density profile relative to the ordinary Newtonian soliton. Fully dynamical pseudospectral simulations with homogeneous and isotropic initial conditions demonstrate that Yukawa screening systematically delays Bose-star condensation, in good agreement with the screened kinetic prediction after fitting a single overall normalization parameter.
\end{abstract}
\maketitle

\section{Introduction}\label{sec:intro}

The nature of dark matter (DM) remains one of the most important unresolved problems in modern cosmology. Observations indicate that DM contributes about $27\%$ of the total energy density of the Universe~\cite{Aghanim:2018eyx}, and many particle candidates have been proposed. Among them, light bosonic dark matter, including QCD axions and ultralight axionlike particles, has attracted significant attention~\cite{2008LNP...741....3P,weinberg1978,Preskill:1982cy,Kim:1979if,Shifman:1979if,Dine:1982ah,Zhitnitsky:1980tq,hu2000,Marsh:2015xka,Hui:2016ltb,Hlozek:2014lca,Hlozek:2017zzf,luzio2020landscape,Sikivie:2006ni}. In the large occupation-number limit, these bosons can be described by a classical complex scalar field in natural units with $\hbar=c=1$. Due to the interplay between gravity, gradient pressure, and possible self-interactions, the bosonic field can undergo gravitational Bose condensation and form compact solitonic objects known as boson stars or solitons~\cite{1968PhRv..172.1331K,1969PhRv..187.1767R,1991PhRvL..66.1659S,1994PhRvL..72.2516S,PhysRevD.84.043531,Chavanis:2011zm,Amin:2019ums,Eby:2015hsq,PhysRevD.98.023009}.

The formation and evolution of boson stars through kinetic relaxation has been studied extensively over the past decades~\cite{2009PhRvL.103k1301S,2015PhRvD..92j3513G}. In particular, Refs.~\cite{Levkov:2018kau,Kirkpatrick:2020fwd} demonstrated that self-gravitating bosons governed by the Schr\"odinger-Poisson equations can dynamically condense and form Bose stars in the kinetic regime, and derived the corresponding condensation timescale. Subsequent numerical studies investigated the relaxation process in systems with gravity and self-interactions, tested analytic condensation-time models, and followed the formation and growth of boson stars inside dark matter halos~\cite{Schwabe_2016,Veltmaat:2018dfz,Eggemeier:2019jsu,PhysRevD.104.083022,PhysRevD.106.023009,Bar:2018acw}. Short-range self-interactions provide another channel for kinetic relaxation and can interfere with gravitational scattering when the two interactions are present as independent amplitudes~\cite{Jain:2024wpx,Kirkpatrick:2022fwd}. These developments also motivated extensions beyond the minimal scalar scenario, including vector dark matter, polarization dynamics in vector halos, nonminimal gravitational couplings, and the nonlinear dynamics of self-interacting solitons~\cite{PhysRevD.108.083021,PhysRevD.111.043031,Chen_2024,sm3l-z7s3,Zhang:2024bjo}.

In this work, we study how finite-range interactions modify the infrared kinetic relaxation responsible for Bose-star condensation. A Yukawa interaction naturally arises when the mediator of the attractive interaction acquires a finite mass. Similar Yukawa parameterizations frequently appear in phenomenological studies of screened or finite-range gravitational interactions. In the present work, the Yukawa interaction is treated phenomenologically as an effective finite-range modification of the long-range gravitational interaction. This differs from the case in which an additional short-range self-interaction is included together with the long-range interaction. Such an additional interaction could generate interference effects in a more general model. Here, however, we do not include an independent self-interaction channel and focus instead on how the finite interaction range replaces the infrared Coulomb logarithm with a Yukawa transport logarithm over the full range of momentum transfer. In the Schr\"odinger-Newton context, Yukawa screening has previously been introduced to study how finite interaction ranges modify collapse and large-scale density evolution~\cite{Mendonca:2021aeq}. However, the effect of screening on kinetic Bose-star condensation and condensation timescales has not yet been systematically investigated.

The finite interaction range influences the system in two closely related ways. First, it modifies the equilibrium Bose-star profile because the outer region of the bound object no longer experiences a purely Newtonian long-range attraction. Second, it changes the kinetic relaxation process responsible for condensation by regulating small-angle gravitational scattering and replacing the usual Coulomb logarithm with a finite Yukawa transport logarithm. We therefore study both the static YSP soliton solutions and the screened kinetic condensation process within a unified framework.

Using pseudospectral simulations in a periodic box with homogeneous and isotropic initial conditions, we numerically investigate the formation of Bose stars in the YSP system. We compare the measured condensation times with the screened kinetic prediction and determine the overall normalization through numerical fitting.

The remainder of this paper is organized as follows. In Sec.~\ref{sec:YSP}, we derive the YSP equations and introduce the dimensionless form used in the simulations. In Sec.~\ref{sec:static}, we study the static YSP ground-state solutions. In Sec.~\ref{sec:Kinetic}, we derive the screened kinetic condensation timescale by replacing the gravitational Coulomb logarithm with the Yukawa transport logarithm. In Sec.~\ref{sec:Numerical_initial}, we describe the numerical methods, initial conditions, and numerical simulation setup, including projected density maps, radial profiles, maximum-density evolution, and comparisons between theoretical predictions and simulations. Finally, Sec.~\ref{sec:conclusion} summarizes our conclusions and discusses possible future extensions.

\section{YSP System}
\label{sec:YSP}

In this work, we consider a bosonic field ($\psi$) interacting through a Yukawa-screened attractive potential. The dynamics of the system can be derived from the mean-field action
\begin{align}
S=&\int dt\,d^3x
\left[
\frac{i}{2}\left(\psi^*\dot\psi-\psi\dot\psi^*\right)
-\frac{|\nabla\psi|^2}{2m}
\right.
\nonumber\\
&\left.
-m\Phi\left(|\psi|^2-n\right)
-\frac{1}{8\pi G}
\left(
(\nabla\Phi)^2+\mu_Y^2\Phi^2
\right)
\right],
\label{eq:action}
\end{align}
where $m$ is the boson mass, $G$ is Newton's gravitational constant, $\Phi$ is the Yukawa-screened gravitational potential, and $n=\langle |\psi|^2\rangle$ is the mean number density. The subtraction of the homogeneous background density is the standard finite-volume prescription adopted in periodic Schr\"odinger-Poisson simulations~\cite{Levkov:2018kau,PhysRevD.106.023009,Mendonca:2021aeq}.

Varying the action with respect to $\psi^*$ and $\Phi$, we obtain the YSP equations
\begin{align}
i\frac{\partial \psi}{\partial t}
&=
-\frac{1}{2m}\nabla^2\psi
+m\Phi\psi,
\label{eq:YSP1}
\\
(\nabla^2-\mu_Y^2)\Phi
&=
4\pi Gm
\left(
|\psi|^2-n
\right),
\label{eq:YSP2}
\end{align}
where $\mu_Y$ is the Yukawa screening mass. In the limit $\mu_Y\rightarrow0$, Eqs.~(\ref{eq:YSP1}) and (\ref{eq:YSP2}) reduce to the ordinary Schr\"odinger-Poisson equations.

To simplify the equations, we introduce the dimensionless quantities
\begin{align}
x
&=
\widetilde{x}/(mv_0),
&
t
&=
\widetilde{t}/(mv_0^2),
&
\Phi
&=
\widetilde{\Phi}v_0^2,
\nonumber\\
\psi
&=
\widetilde{\psi}
v_0^2
\sqrt{\frac{m}{4\pi G}},
&
\mu_Y
&=
\widetilde{\mu}_Y mv_0,
&
n
&=
\widetilde{n}
\frac{mv_0^4}{4\pi G},
\label{eq:rescaling}
\end{align}
where $v_0$ is a reference velocity characterizing the initial state. The corresponding interaction range is $\mu_Y^{-1}$. For ultralight dark matter with $m\sim10^{-22}\,\mathrm{eV}$ and $v_0\sim10^{-3}$, the range $\widetilde{\mu}_Y\in[0,1]$ corresponds to mediator masses up to $\mu_Y\sim10^{-25}\,\mathrm{eV}$, as may arise from an ultralight dark-sector mediator. Using these definitions, the dimensionless YSP equations become
\begin{align}
i\frac{\partial}{\partial\widetilde{t}}
\widetilde{\psi}
&=
-\frac{1}{2}
\widetilde{\nabla}^2
\widetilde{\psi}
+
\widetilde{\Phi}
\widetilde{\psi},
\label{eq:YSP1_dim}
\\
\left(
\widetilde{\nabla}^2
-
\widetilde{\mu}_Y^2
\right)
\widetilde{\Phi}
&=
|\widetilde{\psi}|^2-\widetilde{n}.
\label{eq:YSP2_dim}
\end{align}
The dimensionless particle number satisfies
\begin{equation}
\widetilde N
=
\int d^3\widetilde x\,
|\widetilde\psi|^2
=
4\pi \widetilde M .
\label{eq:Ntilde}
\end{equation}

\section{Static YSP Bose Stars}
\label{sec:static}
\begin{figure}[t]
\centering
\includegraphics[width=\columnwidth]{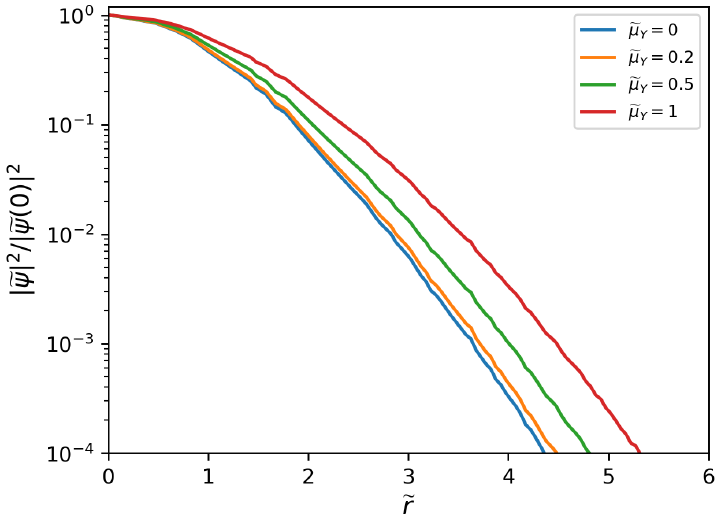}
\caption{
Static YSP Bose-star profiles obtained by imaginary-time relaxation for different Yukawa masses $\widetilde{\mu}_Y$. 
The profiles are normalized to the same central density to illustrate how Yukawa screening modifies the soliton structure. 
The case $\widetilde{\mu}_Y=0$ corresponds to the ordinary Schr\"odinger-Poisson soliton.
}
\label{fig:static_profile}
\end{figure}

We first consider static solutions of the time-independent Yukawa-SP equations using the ansatz
\begin{equation}
\widetilde\psi(\widetilde{\bm{x}},\widetilde t)
=
e^{-i\widetilde\omega\widetilde t}\widetilde\psi_s(\widetilde r),
\end{equation}
where $\widetilde r=|\widetilde{\bm{x}}|$. For localized static solutions, the homogeneous background subtraction is omitted. The static YSP equations are then
\begin{align}
\widetilde\omega\widetilde\psi_s
&=
-\frac{1}{2}\widetilde\nabla^2\widetilde\psi_s+\widetilde\Phi_s\widetilde\psi_s,
\label{eq:static_schrodinger}\\
(\widetilde\nabla^2-\widetilde\mu_Y^2)\widetilde\Phi_s
&=
|\widetilde\psi_s|^2 .
\label{eq:static_yukawa}
\end{align}
We solve Eqs.~\eqref{eq:static_schrodinger} and \eqref{eq:static_yukawa} by imaginary-time relaxation. The solutions are computed at fixed total mass, while the plotted profiles are normalized to the same central density. Increasing $\widetilde\mu_Y$ shortens the range of the attractive potential. The screened solitonic solutions become broader, since a shorter-range attraction requires a less compact configuration at fixed mass. At fixed mass this broadening is accompanied by a shallower bound-state potential, qualitatively similar to the way repulsive short-range interactions broaden self-gravitating condensates~\cite{PhysRevD.104.083022}. This effect is illustrated by comparing the ordinary Schr\"odinger-Poisson profile with the screened YSP profiles in Fig.~\ref{fig:static_profile}.

\section{Screened Kinetic Condensation}
\label{sec:Kinetic}
The condensation time of Bose stars in the kinetic regime is expected to be proportional to the relaxation time of the bosonic system~\cite{Levkov:2018kau,PhysRevD.106.023009}. For ordinary gravitational interactions, the condensation time takes the form
\begin{equation}
\tau_{\rm gr}
=
\frac{b_{\rm gr}\sqrt{2}}{12\pi^3}
\frac{mv^6}{G^2n^2\ln(mvR)},
\label{eq:tau_gr}
\end{equation}
where $v$ is the characteristic velocity dispersion, $n$ is the average number density, $R$ is the size of the virialized region, and $b_{\rm gr}$ is an $\mathcal{O}(1)$ coefficient depending on the initial distribution. For the Yukawa-screened interaction, the transport cross section is modified by the finite interaction range. Using the screened Rutherford scattering kernel with infrared cutoff 
$q_{\rm min}=R^{-1}$ and ultraviolet cutoff $q_{\rm max}=mv$, 
following the standard kinetic treatment of gravitational condensation~\cite{Levkov:2018kau,PhysRevD.106.023009}, 
we obtain the screened transport logarithm. In the comparison with simulations, $R$ is identified with the simulation box size $\widetilde L$ and is held fixed as $\widetilde{\mu}_Y$ is varied, while $v$ is the characteristic initial velocity $v_0$.
\begin{equation}
\Lambda_Y
=
\frac{1}{2}
\left[
\ln
\frac{m^2v^2+\mu_Y^2}
{R^{-2}+\mu_Y^2}
+
\frac{\mu_Y^2}{m^2v^2+\mu_Y^2}
-
\frac{\mu_Y^2}{R^{-2}+\mu_Y^2}
\right].
\label{eq:lambdaY}
\end{equation}
The corresponding screened condensation time becomes
\begin{equation}
\tau_Y
=
\frac{b\sqrt{2}}{6\pi^3}
\frac{mv^6}{G^2n^2\Lambda_Y}.
\label{eq:tauY}
\end{equation}

The numerical prefactor in Eq.~\eqref{eq:tauY} follows from the transport cross section derived in Appendix~\ref{app:derivation}, while the remaining order-unity normalization is absorbed into the fitted coefficient $b$. Equation~\eqref{eq:tauY} has the same logarithmic dependence as the ordinary gravitational result in the limit $\mu_Y\rightarrow0$. As the Yukawa screening mass increases, the long-range part of the interaction becomes suppressed, reducing the transport logarithm $\Lambda_Y$ and increasing the condensation time relative to the Newtonian case. Physically, Yukawa screening suppresses infrared momentum transfer and therefore weakens the kinetic relaxation responsible for Bose-star condensation. Since order-unity factors depend on the initial momentum distribution and on the criterion used to identify condensation, the simulations below mainly test the predicted dependence on $\Lambda_Y$.

\section{Numerical Simulations}
\label{sec:Numerical_initial}
To solve Eqs.~(\ref{eq:YSP1_dim}) and (\ref{eq:YSP2_dim}), we employ a fourth-order pseudospectral method similar to previous Schr\"odinger-Poisson simulations of wave dark matter~\cite{Widrow:1993qq,Schive:2014dra,Mocz:2017wlg,Levkov:2018kau,PhysRevD.106.023009}.
The screened gravitational potential is computed in Fourier space at each time step.

For the initial conditions, we consider a Dirac-delta momentum-shell distribution,
$|\psi_{\bm{p}}|^2
= N\delta(|\bm{p}|-mv_0)$
following the setup of Refs.~\cite{Levkov:2018kau,PhysRevD.104.083022,PhysRevD.108.083021}. The simulations are carried out in a periodic box with dimensionless size $\widetilde{L}$, where the total number of nonrelativistic bosons is given by $N\equiv nL^3$. The initial field configuration in position space, $\psi(\vec x,0)$, is obtained by performing an inverse Fourier transform of $|\psi_{\bm{p}}|e^{iS}$ with randomly distributed phases $S$. This construction produces an initially homogeneous and isotropic random-wave configuration. To investigate the effect of Yukawa screening, we vary the dimensionless screening parameter $\widetilde{\mu}_Y$ within the range $[0,1]$.

We consider simulation box sizes in the range $30\leq\widetilde{L}\leq50$, total mass parameters $15\leq\widetilde{M}\leq100$, and Yukawa screening parameters $0\leq\widetilde{\mu}_Y\leq1.0$. These parameters are chosen such that Bose-star formation occurs within the accessible simulation time interval while allowing comparisons between different screening strengths and system sizes.

To verify that the condensed object is indeed a Bose star, we also compare the density profile measured from simulations with the corresponding static YSP ground-state solution. The static profile is obtained numerically using imaginary-time relaxation.

\begin{figure}[t]
\centering
\includegraphics[width=\columnwidth]{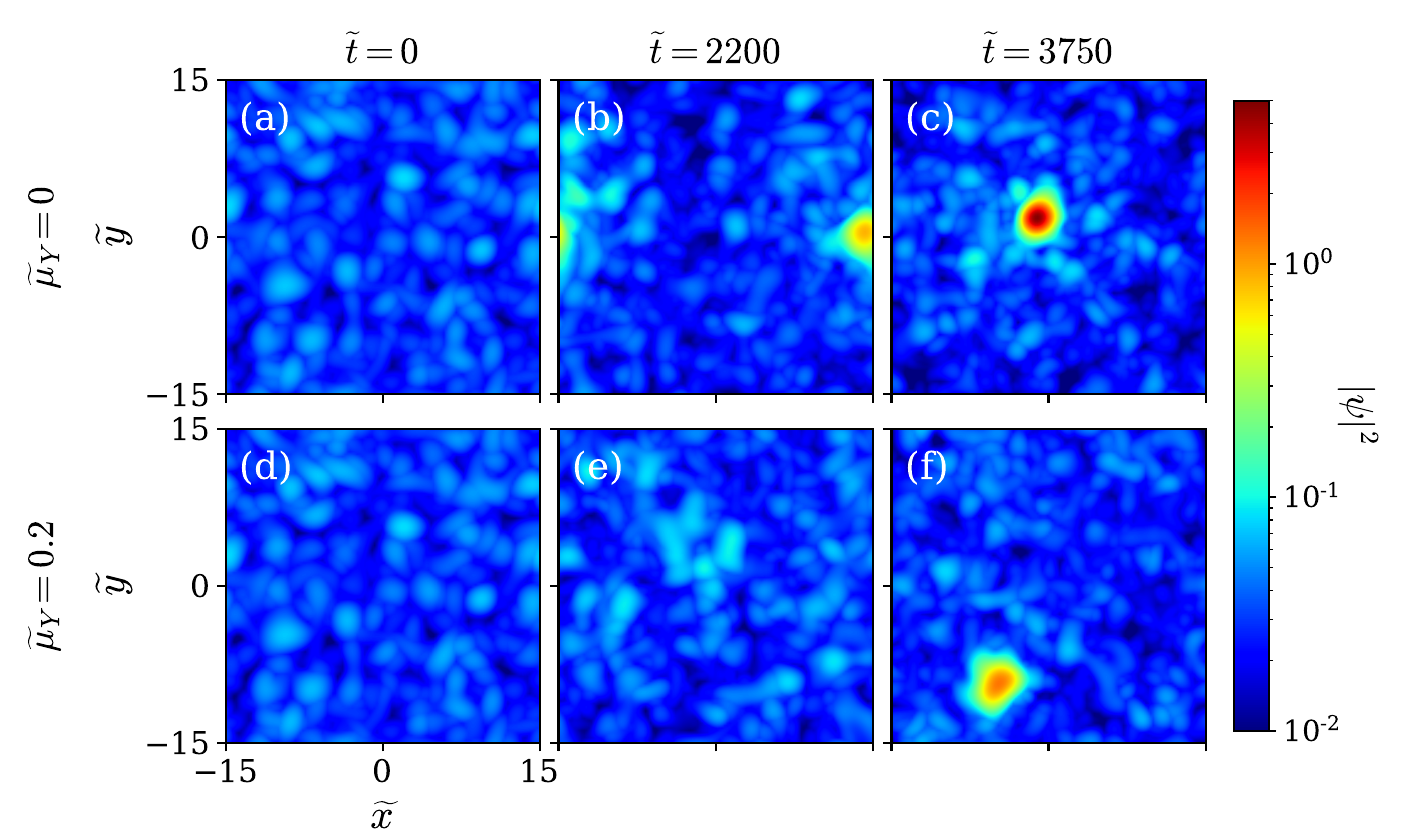}
\caption{Snapshots of the projected density field for simulations with $\widetilde{L}=30$ and $\widetilde{M} = 20$. The top row corresponds to the ordinary Schr\"odinger-Poisson case with $\widetilde{\mu}_Y=0$, while the bottom row shows the Yukawa-screened evolution with $\widetilde{\mu}_Y=0.20$. From left to right, the snapshots are taken at $\widetilde{t}=0$, $\widetilde{t}=2200$, and $\widetilde{t}=3750$, respectively.}
\label{fig:projection}
\end{figure}

Figure~\ref{fig:projection} presents representative snapshots of the projected density field for simulations with $\widetilde{L}=30$ and $\widetilde{M}= 20$. The upper panels correspond to the ordinary Schr\"odinger-Poisson case with $\widetilde{\mu}_Y=0$, while the lower panels show the Yukawa-screened case with $\widetilde{\mu}_Y=0.20$. We can see at $\widetilde{t}=0$, both systems exhibit nearly homogeneous and isotropic random-wave configurations. At the intermediate time, $\widetilde{t} =2200$, the unscreened system, $\widetilde{\mu}_Y=0$ , already develops a prominent compact overdensity associated with Bose-star formation, whereas the screened system, $\widetilde{\mu}_Y=0.20$, remains comparatively diffuse. At the later time, $\widetilde{t} =3750$, a compact Bose star also forms in the Yukawa-screened case, although with a broader structure and delayed condensation compared with the ordinary SP evolution.

\begin{figure}[t]
\centering
\includegraphics[width=\columnwidth]{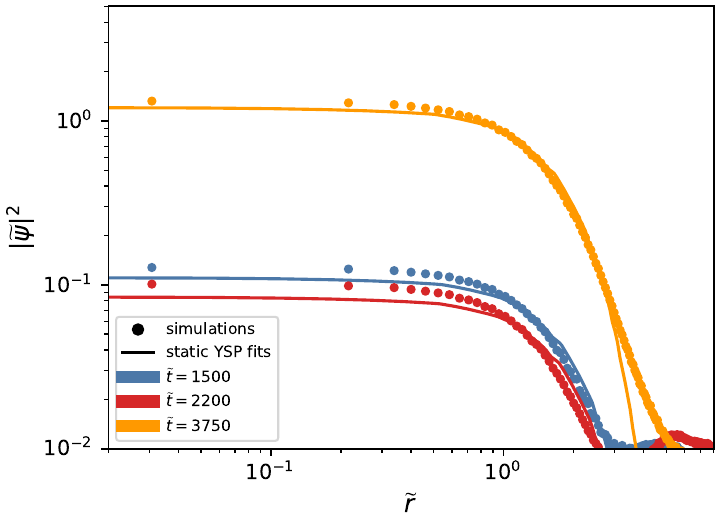}
\caption{Spherically averaged density profiles for the simulation with $\widetilde{L}=30$, $\widetilde{M} = 20$, and $\widetilde{\mu}_Y=0.20$. Colored markers show density profiles extracted from simulations at different evolution times, $\widetilde{t}=1500$, $2200$, and $3750$. Solid curves denote the corresponding fitted static YSP ground-state solutions.}
\label{fig:profile}
\end{figure}

To verify the solitonic nature of the condensed object, we compare representative spherically averaged density profiles extracted from the simulations with static YSP ground-state solutions in Fig.~\ref{fig:profile}. At early time, represented here by the snapshot at $\widetilde{t}=1500$ and $2200$, the density distribution remains relatively diffuse and does not exhibit a well-defined solitonic core. As the system evolves, a compact overdensity gradually emerges through nonlinear relaxation. At later times, $\widetilde{t} = 3750$, the central density profiles are well described by the corresponding screened ground-state solutions, indicating the formation and subsequent growth of a Bose star in the Yukawa-screened system.

\begin{figure}[t]
\centering
\includegraphics[width=\columnwidth]{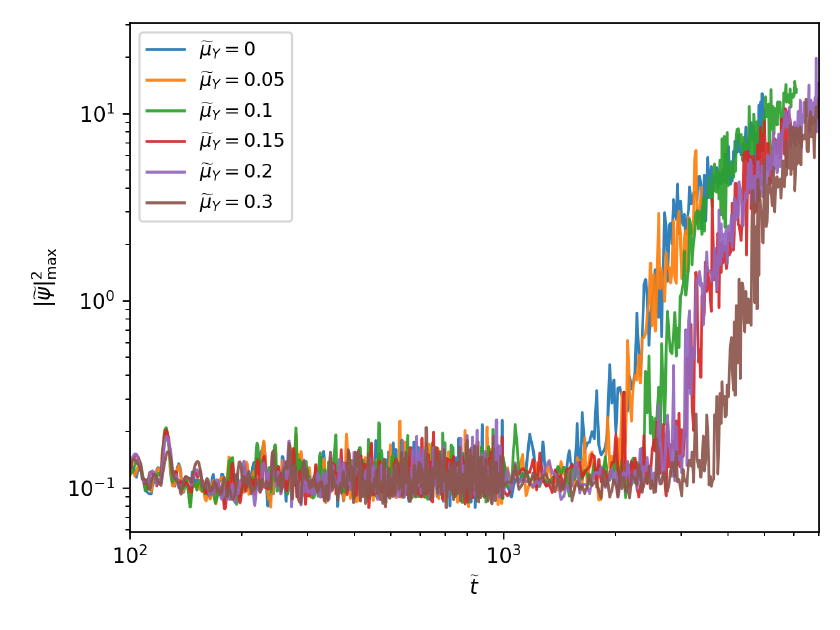}
\caption{Evolution of the maximum density for simulations with box size $\widetilde{L}=30$, total mass $\widetilde{M} = 20$, and different Yukawa screening mass $\widetilde{\mu}_Y$.}
\label{fig:growth}
\end{figure}

\begin{figure}[t]
\centering
\includegraphics[width=\columnwidth]{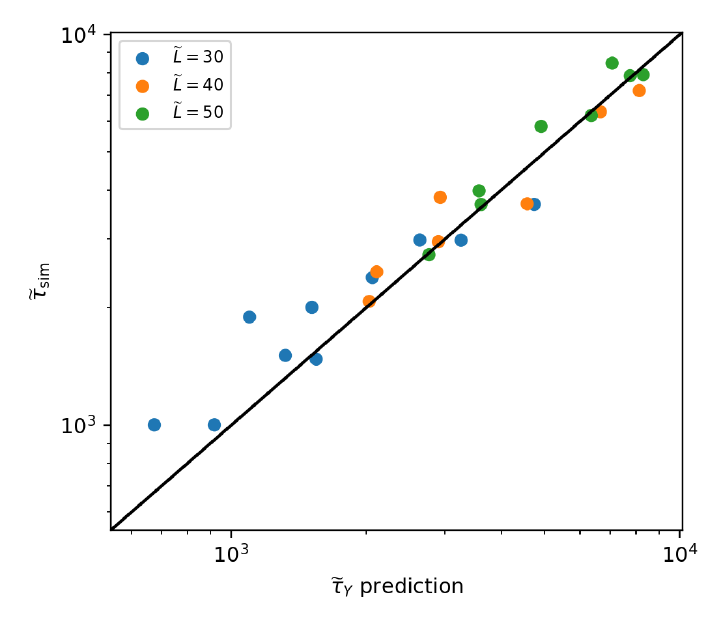}
\caption{Comparison between the predicted condensation times and the measured simulation results for Yukawa-screened runs. The prediction is computed using Eq.~\ref{eq:tauY} with a single fitted coefficient. The solid black line denotes perfect agreement. Different colors correspond to different box sizes $\widetilde{L}$.}
\label{fig:parity}
\end{figure}

Figure~\ref{fig:growth} presents the evolution of the maximum density for simulations with different screening masses $\widetilde{\mu}_Y$. We find that increasing the Yukawa screening mass systematically delays the onset of Bose-star condensation, indicating that finite-range interactions suppress the kinetic relaxation process responsible for condensation. The condensation time is defined as the onset of a stable solitonic core, identified by fitting the density profile around the maximum-density peak to the corresponding static YSP ground state. To investigate this effect quantitatively, the corresponding Bose-star condensation times are compared with the kinetic prediction of Eq.~(\ref{eq:tauY}) in Fig.~\ref{fig:parity}. We keep the analytic scaling fixed and fit only a single coefficient, obtaining the best-fit value
\begin{equation}
b_{\rm fit} \approx 0.27.
\label{eq:bfit}
\end{equation}
The agreement in Fig.~\ref{fig:parity} shows that the measured condensation times follow the predicted $\Lambda_Y^{-1}$ dependence across the simulated screening masses and box sizes.

\section{Conclusions}
\label{sec:conclusion}

We have studied Bose-star formation in a YSP system using kinetic theory and fully dynamical pseudospectral simulations. A finite interaction range modifies both the equilibrium structure of the condensed object and the relaxation process responsible for Bose-star formation.

For the static solutions, we find that Yukawa screening broadens the Bose-star density profile relative to the ordinary Schr\"odinger-Poisson soliton by suppressing the long-range part of the attractive interaction. For the kinetic condensation process, we derive a screened condensation-time formula in which the ordinary gravitational Coulomb logarithm is replaced by a finite Yukawa transport logarithm. Fully dynamical simulations show that increasing the Yukawa screening mass systematically delays Bose-star condensation, with the measured condensation times following the predicted $\Lambda_Y^{-1}$ scaling.

Our results suggest that finite-range interactions can significantly modify both the formation timescale and internal structure of self-gravitating bosonic condensates. Beyond the condensation stage, Yukawa screening may also affect the subsequent nonlinear evolution of the compact object, including accretion, oscillations, mergers, and collapse~\cite{Eby:2015hsq,PhysRevD.104.083022,Chen_2024,sm3l-z7s3}. It will therefore be interesting to extend the present framework to cosmological simulations and to vector dark matter systems, where polarization effects may further alter the condensation dynamics and soliton structure~\cite{PhysRevD.108.083021,PhysRevD.111.043031}.

\appendix

\section{Derivation of the Yukawa Condensation Time}
\label{app:derivation}

For two nonrelativistic particles of mass $m$, the Newtonian potential is replaced by the Yukawa-screened potential
\begin{equation}
V_Y(r)=-\frac{Gm^2}{r}e^{-\mu_Y r}.
\label{eq:app_yukawa_potential}
\end{equation}
Its Fourier transform is
\begin{equation}
\widetilde V_Y(q)
=
-\frac{4\pi Gm^2}{q^2+\mu_Y^2}.
\label{eq:app_vq}
\end{equation}
The scattering is dominated by small deflections.  In this limit one may either use the Born amplitude or, equivalently, the screened Rutherford form.  The momentum transfer is
\begin{equation}
q=2mv\sin\frac{\theta}{2}\simeq mv\theta ,
\label{eq:app_qtheta}
\end{equation}
where $v$ is the relative velocity and $\theta$ is the scattering angle.  The Yukawa mass cuts off the Rutherford singularity at angles of order
\begin{equation}
\theta_Y=\frac{\mu_Y}{mv}.
\label{eq:app_theta_y_def}
\end{equation}
Using the same small-angle treatment as in the gravitational kinetic calculation, we obtain
\begin{equation}
\frac{d\sigma_Y}{d\Omega}
=
\frac{4G^2m^2}{v^4}
\frac{1}{\left(\theta^2+\theta_Y^2\right)^2},
\qquad
\theta_Y\equiv\frac{\mu_Y}{mv},
\label{eq:app_dsigma}
\end{equation}
which reduces to the usual gravitational Rutherford expression for $\mu_Y\rightarrow0$.

The total cross section is not the relevant quantity for relaxation, because very small deflections barely change the particle momentum.  The relaxation rate is controlled by the transport cross section,
\begin{equation}
\sigma_{\rm tr,Y}
=
\int d\Omega\,(1-\cos\theta)
\frac{d\sigma_Y}{d\Omega}.
\label{eq:app_sigtr_def}
\end{equation}
Using $1-\cos\theta\simeq\theta^2/2$, $d\Omega\simeq2\pi\theta\,d\theta$, and $q=mv\theta$, Eq.~\eqref{eq:app_sigtr_def} becomes
\begin{align}
\sigma_{\rm tr,Y}
&=
\int 2\pi\theta\,d\theta\,
\frac{\theta^2}{2}
\frac{4G^2m^2}{v^4}
\frac{1}{(\theta^2+\theta_Y^2)^2}
\nonumber\\
&=
\frac{4\pi G^2m^2}{v^4}
\int_{\theta_{\rm min}}^{\theta_{\rm max}}
\frac{\theta^3\,d\theta}{(\theta^2+\theta_Y^2)^2}.
\label{eq:app_sigtr_theta}
\end{align}
Changing variables from angle to momentum transfer, $q=mv\theta$, gives $d\theta=dq/(mv)$ and
\begin{equation}
\frac{\theta^3\,d\theta}{(\theta^2+\theta_Y^2)^2}
=
\frac{q^3\,dq}{(q^2+\mu_Y^2)^2}.
\label{eq:app_angle_to_q}
\end{equation}
This gives the same screened transport logarithm that replaces the Coulomb logarithm. Using the same transport-cross-section definition as in Eq.~\eqref{eq:app_sigtr_def}, we write
\begin{equation}
\sigma_{\rm tr,Y}
=
\frac{4\pi G^2m^2}{v^4}\Lambda_Y,
\label{eq:app_sigtr}
\end{equation}
where
\begin{equation}
\Lambda_Y
=
\int_{q_{\rm min}}^{q_{\rm max}}
\frac{q^3\,dq}{(q^2+\mu_Y^2)^2}.
\label{eq:app_lambda_integral}
\end{equation}
The lower cutoff is fixed by the finite size of the system,
\begin{equation}
q_{\rm min}=R^{-1},
\label{eq:app_qmin}
\end{equation}
because modes with wavelengths larger than the system do not contribute to kinetic relaxation. In the comparison with simulations we identify $R$ with the box size $\widetilde L$ and keep it fixed as $\widetilde{\mu}_Y$ is varied. The upper cutoff is fixed by the typical momentum transfer in the gas,
\begin{equation}
q_{\rm max}=mv .
\label{eq:app_qmax}
\end{equation}
To evaluate Eq.~\eqref{eq:app_lambda_integral}, set $u=q^2+\mu_Y^2$.  Then $du=2q\,dq$ and $q^2=u-\mu_Y^2$, so
\begin{align}
\int \frac{q^3\,dq}{(q^2+\mu_Y^2)^2}
&=
\frac{1}{2}\int \frac{u-\mu_Y^2}{u^2}\,du
\nonumber\\
&=
\frac{1}{2}
\left[
\ln u+\frac{\mu_Y^2}{u}
\right].
\label{eq:app_lambda_indef}
\end{align}
Evaluating between $q_{\rm min}$ and $q_{\rm max}$ gives
\begin{align}
\Lambda_Y
&=
\frac{1}{2}
\left[
\ln(q^2+\mu_Y^2)
+
\frac{\mu_Y^2}{q^2+\mu_Y^2}
\right]_{q_{\rm min}}^{q_{\rm max}}
\nonumber\\
&=
\frac{1}{2}
\left[
\ln\frac{m^2v^2+\mu_Y^2}{R^{-2}+\mu_Y^2}
+
\frac{\mu_Y^2}{m^2v^2+\mu_Y^2}
-
\frac{\mu_Y^2}{R^{-2}+\mu_Y^2}
\right],
\label{eq:app_lambda_result}
\end{align}
which is Eq.~\eqref{eq:lambdaY}.  In the unscreened limit $\mu_Y\rightarrow0$ this reduces to
\begin{equation}
\Lambda_Y \rightarrow \ln(mvR).
\label{eq:app_newton_limit}
\end{equation}

The screened kinetic condensation time is proportional to the relaxation time of a highly occupied bosonic gas.  A classical wave mode with occupation number $f\gg1$ has a Bose-enhanced scattering rate: the ordinary two-body rate $\sigma_{\rm tr}vn$ is multiplied by the typical final-state occupation.  Following Refs.~\cite{Levkov:2018kau,PhysRevD.106.023009}, the condensation time can therefore be written as
\begin{equation}
\tau
=
\frac{4b\sqrt{2}}{\sigma_{\rm tr} v n f},
\label{eq:app_tau_relax}
\end{equation}
where $b=O(1)$ depends on the initial momentum distribution and on the criterion used to identify condensation.  The factor $4\sqrt{2}$ follows the convention used in the kinetic simulations of gravitational condensation. For the delta-shell initial conditions, residual normalization effects are absorbed into $b$. For an isotropic distribution with characteristic momentum width $mv$, the characteristic phase-space occupation is
\begin{equation}
f=\frac{6\pi^2 n}{(mv)^3}
\label{eq:app_phase_space}
\end{equation}
where $n$ is the number density.  The particles occupy a momentum-space volume of order $(mv)^3$ in the standard phase-space measure. Using Eqs.~\eqref{eq:app_sigtr} and \eqref{eq:app_phase_space} fixes the dependence on $m$, $v$, $n$, and $\Lambda_Y$. We then obtain
\begin{align}
\tau_Y
&=
\frac{4b\sqrt{2}}
{\left(4\pi G^2m^2\Lambda_Y/v^4\right) v n \left[6\pi^2 n/(mv)^3\right]}
\nonumber\\
&=
\frac{b\sqrt{2}}{6\pi^3}
\frac{m v^6}{G^2n^2\Lambda_Y}.
\label{eq:app_tau_y}
\end{align}
This is Eq.~\eqref{eq:tauY}, with $b$ determined from the simulations. The normalization convention used here differs by an overall factor of two from that underlying the gravitational coefficient $b_{\rm gr}$ in Eq.~\eqref{eq:tau_gr}; therefore the fitted $b$ should not be compared directly with $b_{\rm gr}$. The simulations primarily test the predicted $\Lambda_Y^{-1}$ dependence.
\begin{acknowledgments}
We would especially like to thank Zhipan Li for assistance in setting up the clusters. We thank Jiajie Li for helpful comments and assistance in revising the manuscript. JC acknowledges the support from the Fundamental Research Funds for the Central Universities (SWU-KR22012) and the Chongqing Natural Science Foundation General Project (CSTB2023NSCQ-MSX0453). This research was also facilitated by the computational resources in the School of Physical Science and Technology at Southwest University.
\end{acknowledgments}

\bibliographystyle{apsrev4-2}
\bibliography{references}

\begin{thebibliography}{45}%
\makeatletter
\providecommand \@ifxundefined [1]{%
 \@ifx{#1\undefined}
}%
\providecommand \@ifnum [1]{%
 \ifnum #1\expandafter \@firstoftwo
 \else \expandafter \@secondoftwo
 \fi
}%
\providecommand \@ifx [1]{%
 \ifx #1\expandafter \@firstoftwo
 \else \expandafter \@secondoftwo
 \fi
}%
\providecommand \natexlab [1]{#1}%
\providecommand \enquote  [1]{``#1''}%
\providecommand \bibnamefont  [1]{#1}%
\providecommand \bibfnamefont [1]{#1}%
\providecommand \citenamefont [1]{#1}%
\providecommand \href@noop [0]{\@secondoftwo}%
\providecommand \href [0]{\begingroup \@sanitize@url \@href}%
\providecommand \@href[1]{\@@startlink{#1}\@@href}%
\providecommand \@@href[1]{\endgroup#1\@@endlink}%
\providecommand \@sanitize@url [0]{\catcode `\\12\catcode `\$12\catcode
  `\&12\catcode `\#12\catcode `\^12\catcode `\_12\catcode `\%12\relax}%
\providecommand \@@startlink[1]{}%
\providecommand \@@endlink[0]{}%
\providecommand \url  [0]{\begingroup\@sanitize@url \@url }%
\providecommand \@url [1]{\endgroup\@href {#1}{\urlprefix }}%
\providecommand \urlprefix  [0]{URL }%
\providecommand \Eprint [0]{\href }%
\providecommand \doibase [0]{https://doi.org/}%
\providecommand \selectlanguage [0]{\@gobble}%
\providecommand \bibinfo  [0]{\@secondoftwo}%
\providecommand \bibfield  [0]{\@secondoftwo}%
\providecommand \translation [1]{[#1]}%
\providecommand \BibitemOpen [0]{}%
\providecommand \bibitemStop [0]{}%
\providecommand \bibitemNoStop [0]{.\EOS\space}%
\providecommand \EOS [0]{\spacefactor3000\relax}%
\providecommand \BibitemShut  [1]{\csname bibitem#1\endcsname}%
\let\auto@bib@innerbib\@empty
\bibitem [{\citenamefont {Aghanim}\ \emph {et~al.}(2020)\citenamefont {Aghanim}
  \emph {et~al.}}]{Aghanim:2018eyx}%
  \BibitemOpen
  \bibfield  {author} {\bibinfo {author} {\bibfnamefont {N.}~\bibnamefont
  {Aghanim}} \emph {et~al.} (\bibinfo {collaboration} {Planck}),\ }\href
  {https://doi.org/10.1051/0004-6361/201833910} {\bibfield  {journal} {\bibinfo
   {journal} {Astron. Astrophys.}\ }\textbf {\bibinfo {volume} {641}},\
  \bibinfo {pages} {A6} (\bibinfo {year} {2020})},\ \Eprint
  {https://arxiv.org/abs/1807.06209} {arXiv:1807.06209 [astro-ph.CO]}
  \BibitemShut {NoStop}%
\bibitem [{\citenamefont {{Peccei}}(2008)}]{2008LNP...741....3P}%
  \BibitemOpen
  \bibfield  {author} {\bibinfo {author} {\bibfnamefont {R.~D.}\ \bibnamefont
  {{Peccei}}},\ }in\ \href@noop {} {\emph {\bibinfo {booktitle} {Axions}}},\
  \bibinfo {series} {Lecture Notes in Physics, Berlin Springer Verlag}, Vol.\
  \bibinfo {volume} {741},\ \bibinfo {editor} {edited by\ \bibinfo {editor}
  {\bibfnamefont {M.}~\bibnamefont {{Kuster}}}, \bibinfo {editor}
  {\bibfnamefont {G.}~\bibnamefont {{Raffelt}}},\ and\ \bibinfo {editor}
  {\bibfnamefont {B.}~\bibnamefont {{Beltr{\'a}n}}}}\ (\bibinfo {year} {2008})\
  p.~\bibinfo {pages} {3},\ \Eprint {https://arxiv.org/abs/hep-ph/0607268}
  {hep-ph/0607268} \BibitemShut {NoStop}%
\bibitem [{\citenamefont {Weinberg}(1978)}]{weinberg1978}%
  \BibitemOpen
  \bibfield  {author} {\bibinfo {author} {\bibfnamefont {S.}~\bibnamefont
  {Weinberg}},\ }\href {https://doi.org/10.1103/PhysRevLett.40.223} {\bibfield
  {journal} {\bibinfo  {journal} {\prl}\ }\textbf {\bibinfo {volume} {40}},\
  \bibinfo {pages} {223} (\bibinfo {year} {1978})}\BibitemShut {NoStop}%
\bibitem [{\citenamefont {Preskill}\ \emph {et~al.}(1983)\citenamefont
  {Preskill}, \citenamefont {Wise},\ and\ \citenamefont
  {Wilczek}}]{Preskill:1982cy}%
  \BibitemOpen
  \bibfield  {author} {\bibinfo {author} {\bibfnamefont {J.}~\bibnamefont
  {Preskill}}, \bibinfo {author} {\bibfnamefont {M.~B.}\ \bibnamefont {Wise}},\
  and\ \bibinfo {author} {\bibfnamefont {F.}~\bibnamefont {Wilczek}},\ }\href
  {https://doi.org/10.1016/0370-2693(83)90637-8} {\bibfield  {journal}
  {\bibinfo  {journal} {Phys. Lett. B}\ }\textbf {\bibinfo {volume} {120}},\
  \bibinfo {pages} {127} (\bibinfo {year} {1983})}\BibitemShut {NoStop}%
\bibitem [{\citenamefont {Kim}(1979)}]{Kim:1979if}%
  \BibitemOpen
  \bibfield  {author} {\bibinfo {author} {\bibfnamefont {J.~E.}\ \bibnamefont
  {Kim}},\ }\href {https://doi.org/10.1103/PhysRevLett.43.103} {\bibfield
  {journal} {\bibinfo  {journal} {Phys. Rev. Lett.}\ }\textbf {\bibinfo
  {volume} {43}},\ \bibinfo {pages} {103} (\bibinfo {year} {1979})}\BibitemShut
  {NoStop}%
\bibitem [{\citenamefont {Shifman}\ \emph {et~al.}(1980)\citenamefont
  {Shifman}, \citenamefont {Vainshtein},\ and\ \citenamefont
  {Zakharov}}]{Shifman:1979if}%
  \BibitemOpen
  \bibfield  {author} {\bibinfo {author} {\bibfnamefont {M.~A.}\ \bibnamefont
  {Shifman}}, \bibinfo {author} {\bibfnamefont {A.}~\bibnamefont
  {Vainshtein}},\ and\ \bibinfo {author} {\bibfnamefont {V.~I.}\ \bibnamefont
  {Zakharov}},\ }\href {https://doi.org/10.1016/0550-3213(80)90209-6}
  {\bibfield  {journal} {\bibinfo  {journal} {Nucl. Phys. B}\ }\textbf
  {\bibinfo {volume} {166}},\ \bibinfo {pages} {493} (\bibinfo {year}
  {1980})}\BibitemShut {NoStop}%
\bibitem [{\citenamefont {Dine}\ and\ \citenamefont
  {Fischler}(1983)}]{Dine:1982ah}%
  \BibitemOpen
  \bibfield  {author} {\bibinfo {author} {\bibfnamefont {M.}~\bibnamefont
  {Dine}}\ and\ \bibinfo {author} {\bibfnamefont {W.}~\bibnamefont
  {Fischler}},\ }\href {https://doi.org/10.1016/0370-2693(83)90639-1}
  {\bibfield  {journal} {\bibinfo  {journal} {Phys. Lett. B}\ }\textbf
  {\bibinfo {volume} {120}},\ \bibinfo {pages} {137} (\bibinfo {year}
  {1983})}\BibitemShut {NoStop}%
\bibitem [{\citenamefont {Zhitnitsky}(1980)}]{Zhitnitsky:1980tq}%
  \BibitemOpen
  \bibfield  {author} {\bibinfo {author} {\bibfnamefont {A.}~\bibnamefont
  {Zhitnitsky}},\ }\href@noop {} {\bibfield  {journal} {\bibinfo  {journal}
  {Sov.J . Nucl. Phys.}\ }\textbf {\bibinfo {volume} {31}},\ \bibinfo {pages}
  {260} (\bibinfo {year} {1980})}\BibitemShut {NoStop}%
\bibitem [{\citenamefont {Hu}\ \emph {et~al.}(2000)\citenamefont {Hu},
  \citenamefont {Barkana},\ and\ \citenamefont {Gruzinov}}]{hu2000}%
  \BibitemOpen
  \bibfield  {author} {\bibinfo {author} {\bibfnamefont {W.}~\bibnamefont
  {Hu}}, \bibinfo {author} {\bibfnamefont {R.}~\bibnamefont {Barkana}},\ and\
  \bibinfo {author} {\bibfnamefont {A.}~\bibnamefont {Gruzinov}},\ }\href
  {https://doi.org/10.1103/PhysRevLett.85.1158} {\bibfield  {journal} {\bibinfo
   {journal} {\prl}\ }\textbf {\bibinfo {volume} {85}},\ \bibinfo {pages}
  {1158} (\bibinfo {year} {2000})},\ \Eprint
  {https://arxiv.org/abs/astro-ph/0003365} {astro-ph/0003365} \BibitemShut
  {NoStop}%
\bibitem [{\citenamefont {{Marsh}}(2016)}]{Marsh:2015xka}%
  \BibitemOpen
  \bibfield  {author} {\bibinfo {author} {\bibfnamefont {D.~J.~E.}\
  \bibnamefont {{Marsh}}},\ }\href
  {https://doi.org/10.1016/j.physrep.2016.06.005} {\bibfield  {journal}
  {\bibinfo  {journal} {\physrep}\ }\textbf {\bibinfo {volume} {643}},\
  \bibinfo {pages} {1} (\bibinfo {year} {2016})},\ \Eprint
  {https://arxiv.org/abs/1510.07633} {arXiv:1510.07633} \BibitemShut {NoStop}%
\bibitem [{\citenamefont {Hui}\ \emph {et~al.}(2017)\citenamefont {Hui},
  \citenamefont {Ostriker}, \citenamefont {Tremaine},\ and\ \citenamefont
  {Witten}}]{Hui:2016ltb}%
  \BibitemOpen
  \bibfield  {author} {\bibinfo {author} {\bibfnamefont {L.}~\bibnamefont
  {Hui}}, \bibinfo {author} {\bibfnamefont {J.~P.}\ \bibnamefont {Ostriker}},
  \bibinfo {author} {\bibfnamefont {S.}~\bibnamefont {Tremaine}},\ and\
  \bibinfo {author} {\bibfnamefont {E.}~\bibnamefont {Witten}},\ }\href
  {https://doi.org/10.1103/PhysRevD.95.043541} {\bibfield  {journal} {\bibinfo
  {journal} {Phys. Rev.}\ }\textbf {\bibinfo {volume} {D95}},\ \bibinfo {pages}
  {043541} (\bibinfo {year} {2017})},\ \Eprint
  {https://arxiv.org/abs/1610.08297} {arXiv:1610.08297 [astro-ph.CO]}
  \BibitemShut {NoStop}%
\bibitem [{\citenamefont {{Hlozek}}\ \emph {et~al.}(2015)\citenamefont
  {{Hlozek}}, \citenamefont {{Grin}}, \citenamefont {{Marsh}},\ and\
  \citenamefont {{Ferreira}}}]{Hlozek:2014lca}%
  \BibitemOpen
  \bibfield  {author} {\bibinfo {author} {\bibfnamefont {R.}~\bibnamefont
  {{Hlozek}}}, \bibinfo {author} {\bibfnamefont {D.}~\bibnamefont {{Grin}}},
  \bibinfo {author} {\bibfnamefont {D.~J.~E.}\ \bibnamefont {{Marsh}}},\ and\
  \bibinfo {author} {\bibfnamefont {P.~G.}\ \bibnamefont {{Ferreira}}},\
  }\href@noop {} {\bibfield  {journal} {\bibinfo  {journal} {\prd}\ }\textbf
  {\bibinfo {volume} {91}},\ \bibinfo {pages} {103512} (\bibinfo {year}
  {2015})},\ \Eprint {https://arxiv.org/abs/1410.2896} {arXiv:1410.2896}
  \BibitemShut {NoStop}%
\bibitem [{\citenamefont {Hlozek}\ \emph {et~al.}(2018)\citenamefont {Hlozek},
  \citenamefont {Marsh},\ and\ \citenamefont {Grin}}]{Hlozek:2017zzf}%
  \BibitemOpen
  \bibfield  {author} {\bibinfo {author} {\bibfnamefont {R.}~\bibnamefont
  {Hlozek}}, \bibinfo {author} {\bibfnamefont {D.~J.~E.}\ \bibnamefont
  {Marsh}},\ and\ \bibinfo {author} {\bibfnamefont {D.}~\bibnamefont {Grin}},\
  }\href {https://doi.org/10.1093/mnras/sty271} {\bibfield  {journal} {\bibinfo
   {journal} {Mon. Not. Roy. Astron. Soc.}\ }\textbf {\bibinfo {volume}
  {476}},\ \bibinfo {pages} {3063} (\bibinfo {year} {2018})},\ \Eprint
  {https://arxiv.org/abs/1708.05681} {arXiv:1708.05681 [astro-ph.CO]}
  \BibitemShut {NoStop}%
\bibitem [{\citenamefont {Luzio}\ \emph {et~al.}(2020)\citenamefont {Luzio},
  \citenamefont {Giannotti}, \citenamefont {Nardi},\ and\ \citenamefont
  {Visinelli}}]{luzio2020landscape}%
  \BibitemOpen
  \bibfield  {author} {\bibinfo {author} {\bibfnamefont {L.~D.}\ \bibnamefont
  {Luzio}}, \bibinfo {author} {\bibfnamefont {M.}~\bibnamefont {Giannotti}},
  \bibinfo {author} {\bibfnamefont {E.}~\bibnamefont {Nardi}},\ and\ \bibinfo
  {author} {\bibfnamefont {L.}~\bibnamefont {Visinelli}},\ }\href@noop {}
  {\bibinfo {title} {The landscape of qcd axion models}} (\bibinfo {year}
  {2020}),\ \Eprint {https://arxiv.org/abs/2003.01100} {arXiv:2003.01100
  [hep-ph]} \BibitemShut {NoStop}%
\bibitem [{\citenamefont {Sikivie}(2008)}]{Sikivie:2006ni}%
  \BibitemOpen
  \bibfield  {author} {\bibinfo {author} {\bibfnamefont {P.}~\bibnamefont
  {Sikivie}},\ }\href {https://doi.org/10.1007/978-3-540-73518-2_2} {\bibfield
  {journal} {\bibinfo  {journal} {Lect. Notes Phys.}\ }\textbf {\bibinfo
  {volume} {741}},\ \bibinfo {pages} {19} (\bibinfo {year} {2008})},\ \Eprint
  {https://arxiv.org/abs/astro-ph/0610440} {arXiv:astro-ph/0610440}
  \BibitemShut {NoStop}%
\bibitem [{\citenamefont {{Kaup}}(1968)}]{1968PhRv..172.1331K}%
  \BibitemOpen
  \bibfield  {author} {\bibinfo {author} {\bibfnamefont {D.~J.}\ \bibnamefont
  {{Kaup}}},\ }\href {https://doi.org/10.1103/PhysRev.172.1331} {\bibfield
  {journal} {\bibinfo  {journal} {Physical Review}\ }\textbf {\bibinfo {volume}
  {172}},\ \bibinfo {pages} {1331} (\bibinfo {year} {1968})}\BibitemShut
  {NoStop}%
\bibitem [{\citenamefont {{Ruffini}}\ and\ \citenamefont
  {{Bonazzola}}(1969)}]{1969PhRv..187.1767R}%
  \BibitemOpen
  \bibfield  {author} {\bibinfo {author} {\bibfnamefont {R.}~\bibnamefont
  {{Ruffini}}}\ and\ \bibinfo {author} {\bibfnamefont {S.}~\bibnamefont
  {{Bonazzola}}},\ }\href {https://doi.org/10.1103/PhysRev.187.1767} {\bibfield
   {journal} {\bibinfo  {journal} {Physical Review}\ }\textbf {\bibinfo
  {volume} {187}},\ \bibinfo {pages} {1767} (\bibinfo {year}
  {1969})}\BibitemShut {NoStop}%
\bibitem [{\citenamefont {{Seidel}}\ and\ \citenamefont
  {{Suen}}(1991)}]{1991PhRvL..66.1659S}%
  \BibitemOpen
  \bibfield  {author} {\bibinfo {author} {\bibfnamefont {E.}~\bibnamefont
  {{Seidel}}}\ and\ \bibinfo {author} {\bibfnamefont {W.-M.}\ \bibnamefont
  {{Suen}}},\ }\href {https://doi.org/10.1103/PhysRevLett.66.1659} {\bibfield
  {journal} {\bibinfo  {journal} {\prl}\ }\textbf {\bibinfo {volume} {66}},\
  \bibinfo {pages} {1659} (\bibinfo {year} {1991})}\BibitemShut {NoStop}%
\bibitem [{\citenamefont {{Seidel}}\ and\ \citenamefont
  {{Suen}}(1994)}]{1994PhRvL..72.2516S}%
  \BibitemOpen
  \bibfield  {author} {\bibinfo {author} {\bibfnamefont {E.}~\bibnamefont
  {{Seidel}}}\ and\ \bibinfo {author} {\bibfnamefont {W.-M.}\ \bibnamefont
  {{Suen}}},\ }\href {https://doi.org/10.1103/PhysRevLett.72.2516} {\bibfield
  {journal} {\bibinfo  {journal} {\prl}\ }\textbf {\bibinfo {volume} {72}},\
  \bibinfo {pages} {2516} (\bibinfo {year} {1994})},\ \Eprint
  {https://arxiv.org/abs/gr-qc/9309015} {gr-qc/9309015} \BibitemShut {NoStop}%
\bibitem [{\citenamefont {Chavanis}(2011)}]{PhysRevD.84.043531}%
  \BibitemOpen
  \bibfield  {author} {\bibinfo {author} {\bibfnamefont {P.-H.}\ \bibnamefont
  {Chavanis}},\ }\href {https://doi.org/10.1103/PhysRevD.84.043531} {\bibfield
  {journal} {\bibinfo  {journal} {Phys. Rev. D}\ }\textbf {\bibinfo {volume}
  {84}},\ \bibinfo {pages} {043531} (\bibinfo {year} {2011})}\BibitemShut
  {NoStop}%
\bibitem [{\citenamefont {Chavanis}\ and\ \citenamefont
  {Delfini}(2011)}]{Chavanis:2011zm}%
  \BibitemOpen
  \bibfield  {author} {\bibinfo {author} {\bibfnamefont {P.~H.}\ \bibnamefont
  {Chavanis}}\ and\ \bibinfo {author} {\bibfnamefont {L.}~\bibnamefont
  {Delfini}},\ }\href {https://doi.org/10.1103/PhysRevD.84.043532} {\bibfield
  {journal} {\bibinfo  {journal} {Phys. Rev. D}\ }\textbf {\bibinfo {volume}
  {84}},\ \bibinfo {pages} {043532} (\bibinfo {year} {2011})},\ \Eprint
  {https://arxiv.org/abs/1103.2054} {arXiv:1103.2054 [astro-ph.CO]}
  \BibitemShut {NoStop}%
\bibitem [{\citenamefont {Amin}\ and\ \citenamefont
  {Mocz}(2019)}]{Amin:2019ums}%
  \BibitemOpen
  \bibfield  {author} {\bibinfo {author} {\bibfnamefont {M.~A.}\ \bibnamefont
  {Amin}}\ and\ \bibinfo {author} {\bibfnamefont {P.}~\bibnamefont {Mocz}},\
  }\href {https://doi.org/10.1103/PhysRevD.100.063507} {\bibfield  {journal}
  {\bibinfo  {journal} {Phys. Rev. D}\ }\textbf {\bibinfo {volume} {100}},\
  \bibinfo {pages} {063507} (\bibinfo {year} {2019})},\ \Eprint
  {https://arxiv.org/abs/1902.07261} {arXiv:1902.07261 [astro-ph.CO]}
  \BibitemShut {NoStop}%
\bibitem [{\citenamefont {Eby}\ \emph {et~al.}(2016)\citenamefont {Eby},
  \citenamefont {Kouvaris}, \citenamefont {Nielsen},\ and\ \citenamefont
  {Wijewardhana}}]{Eby:2015hsq}%
  \BibitemOpen
  \bibfield  {author} {\bibinfo {author} {\bibfnamefont {J.}~\bibnamefont
  {Eby}}, \bibinfo {author} {\bibfnamefont {C.}~\bibnamefont {Kouvaris}},
  \bibinfo {author} {\bibfnamefont {N.~G.}\ \bibnamefont {Nielsen}},\ and\
  \bibinfo {author} {\bibfnamefont {L.}~\bibnamefont {Wijewardhana}},\ }\href
  {https://doi.org/10.1007/JHEP02(2016)028} {\bibfield  {journal} {\bibinfo
  {journal} {JHEP}\ }\textbf {\bibinfo {volume} {02}},\ \bibinfo {pages}
  {028}},\ \Eprint {https://arxiv.org/abs/1511.04474} {arXiv:1511.04474
  [hep-ph]} \BibitemShut {NoStop}%
\bibitem [{\citenamefont {Chavanis}(2018)}]{PhysRevD.98.023009}%
  \BibitemOpen
  \bibfield  {author} {\bibinfo {author} {\bibfnamefont {P.-H.}\ \bibnamefont
  {Chavanis}},\ }\href {https://doi.org/10.1103/PhysRevD.98.023009} {\bibfield
  {journal} {\bibinfo  {journal} {Phys. Rev. D}\ }\textbf {\bibinfo {volume}
  {98}},\ \bibinfo {pages} {023009} (\bibinfo {year} {2018})}\BibitemShut
  {NoStop}%
\bibitem [{\citenamefont {{Sikivie}}\ and\ \citenamefont
  {{Yang}}(2009)}]{2009PhRvL.103k1301S}%
  \BibitemOpen
  \bibfield  {author} {\bibinfo {author} {\bibfnamefont {P.}~\bibnamefont
  {{Sikivie}}}\ and\ \bibinfo {author} {\bibfnamefont {Q.}~\bibnamefont
  {{Yang}}},\ }\href {https://doi.org/10.1103/PhysRevLett.103.111301}
  {\bibfield  {journal} {\bibinfo  {journal} {\prl}\ }\textbf {\bibinfo
  {volume} {103}},\ \bibinfo {eid} {111301} (\bibinfo {year} {2009})},\ \Eprint
  {https://arxiv.org/abs/0901.1106} {arXiv:0901.1106 [hep-ph]} \BibitemShut
  {NoStop}%
\bibitem [{\citenamefont {{Guth}}\ \emph {et~al.}(2015)\citenamefont {{Guth}},
  \citenamefont {{Hertzberg}},\ and\ \citenamefont
  {{Prescod-Weinstein}}}]{2015PhRvD..92j3513G}%
  \BibitemOpen
  \bibfield  {author} {\bibinfo {author} {\bibfnamefont {A.~H.}\ \bibnamefont
  {{Guth}}}, \bibinfo {author} {\bibfnamefont {M.~P.}\ \bibnamefont
  {{Hertzberg}}},\ and\ \bibinfo {author} {\bibfnamefont {C.}~\bibnamefont
  {{Prescod-Weinstein}}},\ }\href {https://doi.org/10.1103/PhysRevD.92.103513}
  {\bibfield  {journal} {\bibinfo  {journal} {\prd}\ }\textbf {\bibinfo
  {volume} {92}},\ \bibinfo {eid} {103513} (\bibinfo {year} {2015})},\ \Eprint
  {https://arxiv.org/abs/1412.5930} {arXiv:1412.5930} \BibitemShut {NoStop}%
\bibitem [{\citenamefont {Levkov}\ \emph {et~al.}(2018)\citenamefont {Levkov},
  \citenamefont {Panin},\ and\ \citenamefont {Tkachev}}]{Levkov:2018kau}%
  \BibitemOpen
  \bibfield  {author} {\bibinfo {author} {\bibfnamefont {D.~G.}\ \bibnamefont
  {Levkov}}, \bibinfo {author} {\bibfnamefont {A.~G.}\ \bibnamefont {Panin}},\
  and\ \bibinfo {author} {\bibfnamefont {I.~I.}\ \bibnamefont {Tkachev}},\
  }\href {https://doi.org/10.1103/PhysRevLett.121.151301} {\bibfield  {journal}
  {\bibinfo  {journal} {Phys. Rev. Lett.}\ }\textbf {\bibinfo {volume} {121}},\
  \bibinfo {pages} {151301} (\bibinfo {year} {2018})},\ \Eprint
  {https://arxiv.org/abs/1804.05857} {arXiv:1804.05857 [astro-ph.CO]}
  \BibitemShut {NoStop}%
\bibitem [{\citenamefont {Kirkpatrick}\ \emph {et~al.}(2020)\citenamefont
  {Kirkpatrick}, \citenamefont {Mirasola},\ and\ \citenamefont
  {Prescod-Weinstein}}]{Kirkpatrick:2020fwd}%
  \BibitemOpen
  \bibfield  {author} {\bibinfo {author} {\bibfnamefont {K.}~\bibnamefont
  {Kirkpatrick}}, \bibinfo {author} {\bibfnamefont {A.~E.}\ \bibnamefont
  {Mirasola}},\ and\ \bibinfo {author} {\bibfnamefont {C.}~\bibnamefont
  {Prescod-Weinstein}},\ }\href {https://doi.org/10.1103/PhysRevD.102.103012}
  {\bibfield  {journal} {\bibinfo  {journal} {Phys. Rev. D}\ }\textbf {\bibinfo
  {volume} {102}},\ \bibinfo {pages} {103012} (\bibinfo {year} {2020})},\
  \Eprint {https://arxiv.org/abs/2007.07438} {arXiv:2007.07438 [hep-ph]}
  \BibitemShut {NoStop}%
\bibitem [{\citenamefont {Schwabe}\ \emph {et~al.}(2016)\citenamefont
  {Schwabe}, \citenamefont {Niemeyer},\ and\ \citenamefont
  {Engels}}]{Schwabe_2016}%
  \BibitemOpen
  \bibfield  {author} {\bibinfo {author} {\bibfnamefont {B.}~\bibnamefont
  {Schwabe}}, \bibinfo {author} {\bibfnamefont {J.~C.}\ \bibnamefont
  {Niemeyer}},\ and\ \bibinfo {author} {\bibfnamefont {J.~F.}\ \bibnamefont
  {Engels}},\ }\bibfield  {journal} {\bibinfo  {journal} {Physical Review D}\
  }\textbf {\bibinfo {volume} {94}},\ \href
  {https://doi.org/10.1103/physrevd.94.043513} {10.1103/physrevd.94.043513}
  (\bibinfo {year} {2016})\BibitemShut {NoStop}%
\bibitem [{\citenamefont {Veltmaat}\ \emph {et~al.}(2018)\citenamefont
  {Veltmaat}, \citenamefont {Niemeyer},\ and\ \citenamefont
  {Schwabe}}]{Veltmaat:2018dfz}%
  \BibitemOpen
  \bibfield  {author} {\bibinfo {author} {\bibfnamefont {J.}~\bibnamefont
  {Veltmaat}}, \bibinfo {author} {\bibfnamefont {J.~C.}\ \bibnamefont
  {Niemeyer}},\ and\ \bibinfo {author} {\bibfnamefont {B.}~\bibnamefont
  {Schwabe}},\ }\href {https://doi.org/10.1103/PhysRevD.98.043509} {\bibfield
  {journal} {\bibinfo  {journal} {Phys. Rev. D}\ }\textbf {\bibinfo {volume}
  {98}},\ \bibinfo {pages} {043509} (\bibinfo {year} {2018})},\ \Eprint
  {https://arxiv.org/abs/1804.09647} {arXiv:1804.09647 [astro-ph.CO]}
  \BibitemShut {NoStop}%
\bibitem [{\citenamefont {Eggemeier}\ and\ \citenamefont
  {Niemeyer}(2019)}]{Eggemeier:2019jsu}%
  \BibitemOpen
  \bibfield  {author} {\bibinfo {author} {\bibfnamefont {B.}~\bibnamefont
  {Eggemeier}}\ and\ \bibinfo {author} {\bibfnamefont {J.~C.}\ \bibnamefont
  {Niemeyer}},\ }\href {https://doi.org/10.1103/PhysRevD.100.063528} {\bibfield
   {journal} {\bibinfo  {journal} {Phys. Rev. D}\ }\textbf {\bibinfo {volume}
  {100}},\ \bibinfo {pages} {063528} (\bibinfo {year} {2019})},\ \Eprint
  {https://arxiv.org/abs/1906.01348} {arXiv:1906.01348 [astro-ph.CO]}
  \BibitemShut {NoStop}%
\bibitem [{\citenamefont {Chen}\ \emph {et~al.}(2021)\citenamefont {Chen},
  \citenamefont {Du}, \citenamefont {Lentz}, \citenamefont {Marsh},\ and\
  \citenamefont {Niemeyer}}]{PhysRevD.104.083022}%
  \BibitemOpen
  \bibfield  {author} {\bibinfo {author} {\bibfnamefont {J.}~\bibnamefont
  {Chen}}, \bibinfo {author} {\bibfnamefont {X.}~\bibnamefont {Du}}, \bibinfo
  {author} {\bibfnamefont {E.~W.}\ \bibnamefont {Lentz}}, \bibinfo {author}
  {\bibfnamefont {D.~J.~E.}\ \bibnamefont {Marsh}},\ and\ \bibinfo {author}
  {\bibfnamefont {J.~C.}\ \bibnamefont {Niemeyer}},\ }\href
  {https://doi.org/10.1103/PhysRevD.104.083022} {\bibfield  {journal} {\bibinfo
   {journal} {Phys. Rev. D}\ }\textbf {\bibinfo {volume} {104}},\ \bibinfo
  {pages} {083022} (\bibinfo {year} {2021})}\BibitemShut {NoStop}%
\bibitem [{\citenamefont {Chen}\ \emph {et~al.}(2022)\citenamefont {Chen},
  \citenamefont {Du}, \citenamefont {Lentz},\ and\ \citenamefont
  {Marsh}}]{PhysRevD.106.023009}%
  \BibitemOpen
  \bibfield  {author} {\bibinfo {author} {\bibfnamefont {J.}~\bibnamefont
  {Chen}}, \bibinfo {author} {\bibfnamefont {X.}~\bibnamefont {Du}}, \bibinfo
  {author} {\bibfnamefont {E.~W.}\ \bibnamefont {Lentz}},\ and\ \bibinfo
  {author} {\bibfnamefont {D.~J.~E.}\ \bibnamefont {Marsh}},\ }\href
  {https://doi.org/10.1103/PhysRevD.106.023009} {\bibfield  {journal} {\bibinfo
   {journal} {Phys. Rev. D}\ }\textbf {\bibinfo {volume} {106}},\ \bibinfo
  {pages} {023009} (\bibinfo {year} {2022})}\BibitemShut {NoStop}%
\bibitem [{\citenamefont {Bar}\ \emph {et~al.}(2018)\citenamefont {Bar},
  \citenamefont {Blas}, \citenamefont {Blum},\ and\ \citenamefont
  {Sibiryakov}}]{Bar:2018acw}%
  \BibitemOpen
  \bibfield  {author} {\bibinfo {author} {\bibfnamefont {N.}~\bibnamefont
  {Bar}}, \bibinfo {author} {\bibfnamefont {D.}~\bibnamefont {Blas}}, \bibinfo
  {author} {\bibfnamefont {K.}~\bibnamefont {Blum}},\ and\ \bibinfo {author}
  {\bibfnamefont {S.}~\bibnamefont {Sibiryakov}},\ }\href
  {https://doi.org/10.1103/PhysRevD.98.083027} {\bibfield  {journal} {\bibinfo
  {journal} {Phys. Rev. D}\ }\textbf {\bibinfo {volume} {98}},\ \bibinfo
  {pages} {083027} (\bibinfo {year} {2018})},\ \Eprint
  {https://arxiv.org/abs/1805.00122} {arXiv:1805.00122 [astro-ph.CO]}
  \BibitemShut {NoStop}%
\bibitem [{\citenamefont {Jain}\ \emph {et~al.}(2024)\citenamefont {Jain},
  \citenamefont {Wanichwecharungruang},\ and\ \citenamefont
  {Thomas}}]{Jain:2024wpx}%
  \BibitemOpen
  \bibfield  {author} {\bibinfo {author} {\bibfnamefont {M.}~\bibnamefont
  {Jain}}, \bibinfo {author} {\bibfnamefont {W.}~\bibnamefont
  {Wanichwecharungruang}},\ and\ \bibinfo {author} {\bibfnamefont
  {J.}~\bibnamefont {Thomas}},\ }\href
  {https://doi.org/10.1103/PhysRevD.109.016002} {\bibfield  {journal} {\bibinfo
   {journal} {Phys. Rev. D}\ }\textbf {\bibinfo {volume} {109}},\ \bibinfo
  {pages} {016002} (\bibinfo {year} {2024})},\ \Eprint
  {https://arxiv.org/abs/2310.00058} {arXiv:2310.00058 [hep-ph]} \BibitemShut
  {NoStop}%
\bibitem [{\citenamefont {Kirkpatrick}\ \emph {et~al.}(2022)\citenamefont
  {Kirkpatrick}, \citenamefont {Mirasola},\ and\ \citenamefont
  {Prescod-Weinstein}}]{Kirkpatrick:2022fwd}%
  \BibitemOpen
  \bibfield  {author} {\bibinfo {author} {\bibfnamefont {K.}~\bibnamefont
  {Kirkpatrick}}, \bibinfo {author} {\bibfnamefont {A.~E.}\ \bibnamefont
  {Mirasola}},\ and\ \bibinfo {author} {\bibfnamefont {C.}~\bibnamefont
  {Prescod-Weinstein}},\ }\href {https://doi.org/10.1103/PhysRevD.106.043512}
  {\bibfield  {journal} {\bibinfo  {journal} {Phys. Rev. D}\ }\textbf {\bibinfo
  {volume} {106}},\ \bibinfo {pages} {043512} (\bibinfo {year} {2022})},\
  \Eprint {https://arxiv.org/abs/2110.08921} {arXiv:2110.08921 [hep-ph]}
  \BibitemShut {NoStop}%
\bibitem [{\citenamefont {Chen}\ \emph {et~al.}(2023)\citenamefont {Chen},
  \citenamefont {Du}, \citenamefont {Zhou}, \citenamefont {Benson},\ and\
  \citenamefont {Marsh}}]{PhysRevD.108.083021}%
  \BibitemOpen
  \bibfield  {author} {\bibinfo {author} {\bibfnamefont {J.}~\bibnamefont
  {Chen}}, \bibinfo {author} {\bibfnamefont {X.}~\bibnamefont {Du}}, \bibinfo
  {author} {\bibfnamefont {M.}~\bibnamefont {Zhou}}, \bibinfo {author}
  {\bibfnamefont {A.}~\bibnamefont {Benson}},\ and\ \bibinfo {author}
  {\bibfnamefont {D.~J.~E.}\ \bibnamefont {Marsh}},\ }\href@noop {} {\bibfield
  {journal} {\bibinfo  {journal} {Phys. Rev. D}\ }\textbf {\bibinfo {volume}
  {108}},\ \bibinfo {pages} {083021} (\bibinfo {year} {2023})}\BibitemShut
  {NoStop}%
\bibitem [{\citenamefont {Chen}\ \emph {et~al.}(2025)\citenamefont {Chen},
  \citenamefont {Nguyen},\ and\ \citenamefont {Marsh}}]{PhysRevD.111.043031}%
  \BibitemOpen
  \bibfield  {author} {\bibinfo {author} {\bibfnamefont {J.}~\bibnamefont
  {Chen}}, \bibinfo {author} {\bibfnamefont {L.~H.}\ \bibnamefont {Nguyen}},\
  and\ \bibinfo {author} {\bibfnamefont {D.~J.~E.}\ \bibnamefont {Marsh}},\
  }\href@noop {} {\bibfield  {journal} {\bibinfo  {journal} {Phys. Rev. D}\
  }\textbf {\bibinfo {volume} {111}},\ \bibinfo {pages} {043031} (\bibinfo
  {year} {2025})}\BibitemShut {NoStop}%
\bibitem [{\citenamefont {Chen}\ and\ \citenamefont {Zhang}(2024)}]{Chen_2024}%
  \BibitemOpen
  \bibfield  {author} {\bibinfo {author} {\bibfnamefont {J.}~\bibnamefont
  {Chen}}\ and\ \bibinfo {author} {\bibfnamefont {H.-Y.}\ \bibnamefont
  {Zhang}},\ }\href {https://doi.org/10.1088/1475-7516/2024/10/005} {\bibfield
  {journal} {\bibinfo  {journal} {Journal of Cosmology and Astroparticle
  Physics}\ }\textbf {\bibinfo {volume} {2024}}\bibinfo  {number} { (10)},\
  \bibinfo {pages} {005}}\BibitemShut {NoStop}%
\bibitem [{\citenamefont {Zeng}\ \emph {et~al.}(2026)\citenamefont {Zeng},
  \citenamefont {Zhang},\ and\ \citenamefont {Chen}}]{sm3l-z7s3}%
  \BibitemOpen
\bibfield  {number} {  }\bibfield  {author} {\bibinfo {author} {\bibfnamefont
  {Y.}~\bibnamefont {Zeng}}, \bibinfo {author} {\bibfnamefont {B.}~\bibnamefont
  {Zhang}},\ and\ \bibinfo {author} {\bibfnamefont {J.}~\bibnamefont {Chen}},\
  }\href {https://doi.org/10.1103/sm3l-z7s3} {\bibfield  {journal} {\bibinfo
  {journal} {Phys. Rev. D}\ }\textbf {\bibinfo {volume} {113}},\ \bibinfo
  {pages} {103043} (\bibinfo {year} {2026})}\BibitemShut {NoStop}%
\bibitem [{\citenamefont {Zhang}(2025)}]{Zhang:2024bjo}%
  \BibitemOpen
  \bibfield  {author} {\bibinfo {author} {\bibfnamefont {H.-Y.}\ \bibnamefont
  {Zhang}},\ }\href {https://doi.org/10.1007/JHEP04(2025)174} {\bibfield
  {journal} {\bibinfo  {journal} {JHEP}\ }\textbf {\bibinfo {volume} {04}},\
  \bibinfo {pages} {174}},\ \Eprint {https://arxiv.org/abs/2406.05031}
  {arXiv:2406.05031 [hep-ph]} \BibitemShut {NoStop}%
\bibitem [{\citenamefont {Mendon{\c{c}}a}(2021)}]{Mendonca:2021aeq}%
  \BibitemOpen
  \bibfield  {author} {\bibinfo {author} {\bibfnamefont {J.~T.}\ \bibnamefont
  {Mendon{\c{c}}a}},\ }\href {https://doi.org/10.3390/sym13061007} {\bibfield
  {journal} {\bibinfo  {journal} {Symmetry}\ }\textbf {\bibinfo {volume}
  {13}},\ \bibinfo {pages} {1007} (\bibinfo {year} {2021})}\BibitemShut
  {NoStop}%
\bibitem [{\citenamefont {{Widrow}}\ and\ \citenamefont
  {{Kaiser}}(1993)}]{Widrow:1993qq}%
  \BibitemOpen
  \bibfield  {author} {\bibinfo {author} {\bibfnamefont {L.~M.}\ \bibnamefont
  {{Widrow}}}\ and\ \bibinfo {author} {\bibfnamefont {N.}~\bibnamefont
  {{Kaiser}}},\ }\href {https://doi.org/10.1086/187073} {\bibfield  {journal}
  {\bibinfo  {journal} {\apjl}\ }\textbf {\bibinfo {volume} {416}},\ \bibinfo
  {pages} {L71} (\bibinfo {year} {1993})}\BibitemShut {NoStop}%
\bibitem [{\citenamefont {{Schive}}\ \emph {et~al.}(2014)\citenamefont
  {{Schive}}, \citenamefont {{Chiueh}},\ and\ \citenamefont
  {{Broadhurst}}}]{Schive:2014dra}%
  \BibitemOpen
  \bibfield  {author} {\bibinfo {author} {\bibfnamefont {H.-Y.}\ \bibnamefont
  {{Schive}}}, \bibinfo {author} {\bibfnamefont {T.}~\bibnamefont {{Chiueh}}},\
  and\ \bibinfo {author} {\bibfnamefont {T.}~\bibnamefont {{Broadhurst}}},\
  }\href {https://doi.org/10.1038/nphys2996} {\bibfield  {journal} {\bibinfo
  {journal} {Nature Physics}\ }\textbf {\bibinfo {volume} {10}},\ \bibinfo
  {pages} {496} (\bibinfo {year} {2014})},\ \Eprint
  {https://arxiv.org/abs/1406.6586} {arXiv:1406.6586} \BibitemShut {NoStop}%
\bibitem [{\citenamefont {Mocz}\ \emph {et~al.}(2017)\citenamefont {Mocz},
  \citenamefont {Vogelsberger}, \citenamefont {Robles}, \citenamefont {Zavala},
  \citenamefont {Boylan-Kolchin},\ and\ \citenamefont
  {Hernquist}}]{Mocz:2017wlg}%
  \BibitemOpen
  \bibfield  {author} {\bibinfo {author} {\bibfnamefont {P.}~\bibnamefont
  {Mocz}}, \bibinfo {author} {\bibfnamefont {M.}~\bibnamefont {Vogelsberger}},
  \bibinfo {author} {\bibfnamefont {V.}~\bibnamefont {Robles}}, \bibinfo
  {author} {\bibfnamefont {J.}~\bibnamefont {Zavala}}, \bibinfo {author}
  {\bibfnamefont {M.}~\bibnamefont {Boylan-Kolchin}},\ and\ \bibinfo {author}
  {\bibfnamefont {L.}~\bibnamefont {Hernquist}},\ }\href
  {https://doi.org/10.1093/mnras/stx1887} {\bibfield  {journal} {\bibinfo
  {journal} {Mon. Not. Roy. Astron. Soc.}\ }\textbf {\bibinfo {volume} {471}},\
  \bibinfo {pages} {4559} (\bibinfo {year} {2017})},\ \Eprint
  {https://arxiv.org/abs/1705.05845} {arXiv:1705.05845 [astro-ph.CO]}
  \BibitemShut {NoStop}%
\end{thebibliography}%

\end{document}